\documentstyle[12pt,epsf,epsfig]{article}
\def\unit{\hbox to 3.3pt{\hskip1.3pt \vrule height 7pt width .4pt \hskip.7pt
\vrule height 7.85pt width .4pt \kern-2.4pt
\hrulefill \kern-3pt
\raise 4pt\hbox{\char'40}}}

\newcommand{\gtrsim}
{\ \rlap{\raise 2pt\hbox{$>$}}{\lower 2pt \hbox{$\sim$}}\ }
\newcommand{\lessim}
{\ \rlap{\raise 2pt\hbox{$<$}}{\lower 2pt \hbox{$\sim$}}\ }
 
\newcommand{\ea}{{ et al.}}

\def\identity{1 \hspace{-.085cm}{\rm l}}
\def\calo{{\cal O}}

\def\eg{{\it e.g.}}

\def\be{\begin{equation}}
\def\ee{\end{equation}}
\def\bea{\begin{eqnarray}}
\def\eea{\end{eqnarray}}
\def\identity{1 \hspace{-.085cm}{\rm l}}
 
\def\bmu{B_{\mu}} 
\def\beq{\begin{equation}}
\def\eeq{\end{equation}}
\def\bea{\begin{eqnarray}}
\def\eea{\end{eqnarray}}
\def\ba{\begin{array}}
\def\ea{\end{array}}
\def\gappeq{\mathrel{\rlap {\raise.5ex\hbox{$>$}}
{\lower.5ex\hbox{$\sim$}}}}
\def\permil{$\%\raise.20ex\hbox{$_0$}}
\def\lappeq{\mathrel{\rlap{\raise.5ex\hbox{$<$}}
{\lower.5ex\hbox{$\sim$}}}}
\begin{document}
\topmargin -1.0cm
\oddsidemargin -0.8cm
\evensidemargin -0.8cm
\pagestyle{empty}
\begin{flushright}
CERN-TH/96-61
\end{flushright}
\vspace*{5mm}
\begin{center}
{\Large \bf The $\mu$-Problem in Theories with Gauge-Mediated}\\
\vspace{0.5cm}
{\Large\bf Supersymmetry Breaking}\\
\vspace{2.5cm}
{\large G.~Dvali,
G.F.~Giudice\footnote{On leave of absence from INFN, Sezione di
Padova,
Padua, Italy.} and A.~Pomarol}\\
\vspace{0.3cm}
Theory Division, CERN\\
Geneva, Switzerland\\
\vspace*{2cm}
Abstract
\end{center}

We point out that the $\mu$-problem in theories in which supersymmetry 
breaking is communicated to the observable sector by gauge interactions 
is more severe than the one encountered
in the conventional gravity-mediated scenarios.
The difficulty is that once $\mu$ is generated by a one-loop diagram,
then usually $\bmu$ is also generated at the same loop order. This leads 
to the problematic relation $\bmu \sim \mu \Lambda$, where $\Lambda \sim$ 
10--100 TeV is the effective supersymmetry-breaking scale. We present a 
class of theories for which this problem is naturally solved. Here, 
without any fine tuning among parameters, $\mu$ is generated at one loop, 
while $\bmu$ arises only at the two-loop level.
This mechanism can naturally lead to an interpretation of the Higgs
doublets as pseudo-Goldstone bosons of an approximate global symmetry.
 
\vfill
\begin{flushleft}
CERN-TH/96-61\\
March 1996
\end{flushleft}
\eject
\pagestyle{empty}
%\clearpage\mbox{}\clearpage
\setcounter{page}{1}
\setcounter{footnote}{0}

\baselineskip20pt
\pagestyle{plain}

\section{Introduction}

During the early 1980's there was some considerable 
effort in developing theories
with supersymmetry breaking originating in some hidden sector and then 
communicating with the observable sector via gauge interactions at the 
quantum 
level \cite{gaug}. The goal was to construct realistic models which could
circumvent the obstacle imposed by the tree-level mass sum rule in global
supersymmetry \cite{fer}. Since then, these models have largely been
abandoned in favour of the more promising theories where gravity
mediates the supersymmetry breaking \cite{gra}.
However, the puzzle of explaining the suppression of Flavour-Changing
Neutral-Current (FCNC) processes in supergravity 
theories have recently revived
the interest \cite{din,dim} in models with Gauge-Mediated Supersymmetry
Breaking (GMSB). Indeed, in this class of theories, the FCNC problem is 
naturally solved as gauge interactions provide flavour-symmetric
supersymmetry-breaking terms in the observable sector.

In this paper we want to point out that in general GMSB theories suffer 
from a $\mu$-problem. Usually one refers to the $\mu$-problem as 
the difficulty in generating the correct mass scale for the Higgs
bilinear term in the superpotential
\beq
W=\mu {\bar H}H~,
\label{mu}
\eeq
which, for phenomenological reasons, has to be of the order of the weak
scale. 

In supergravity, if the term in eq.~(\ref{mu}) is forbidden in the limit 
of exact supersymmetry, it is then generated at the correct scale as
an effect of supersymmetry breaking, as long as the K\"ahler metric has
not a minimal form \cite{giu}. The $\mu$-problem in GMSB theories 
is, in a certain way, more severe.
Suppose that the term in eq.~(\ref{mu}) is forbidden
in the limit of exact supersymmetry. As we will show in sect.~3,
it is not difficult to envisage
new interactions which generate $\mu$ at one loop, $\mu \sim \frac{
\lambda^2}{(4\pi)^2}\Lambda$; here $\lambda$ is some new coupling constant
and $\Lambda$ is the effective scale of supersymmetry breaking. It is
fairly generic that the same interactions generate also the other
soft supersymmetry-breaking terms in the Higgs potential at the same
one-loop level, $m_H^2 \sim m_{\bar H}^2 \sim B_\mu \sim 
\frac{\lambda^2}{(4\pi)^2}\Lambda^2$, where
\beq
V_{\rm soft}=m_H^2|H|^2+m_{\bar H}^2|{\bar H}|^2 
+(B_\mu {\bar H}H
+{\rm h.c.})~.
\eeq
An attractive feature of GMSB is
that gauginos receive masses at one loop, $m_\lambda \sim \frac{
\alpha}{4\pi}\Lambda$, while squarks and sleptons do so only at two loops,
${\widetilde m}^2 \sim (\frac{\alpha}{4\pi})^2\Lambda^2$. Because of the
different dimensionalities between fermionic and scalar mass terms, this
implies that the gaugino and squark mass scales are of the same order,
$m_{\lambda}\sim {\widetilde m}$.
On the other hand, in the case of the
Higgs parameters, we find $B_\mu \sim \mu \Lambda$, as parameters with
different dimensionalities are generated 
at the same loop level. Since $\Lambda$ 
must be in the range
10--100 TeV to generate appropriate squark and gaugino masses, then
either $\mu$ is at the weak scale and $B_\mu$ violates
the naturalness criterion \cite{bar}, or $B_\mu$ is at the weak scale
and $\mu$ is unacceptably small. We will refer to this puzzle as to
the $\mu$-problem in GMSB theories.

We wish to stress that this is only an ``aesthetic" problem
and not an inconsistency of the theory.
It is certainly possible to introduce new interactions able to generate
separately both $\mu$ and $B_\mu$, but this requires {\it ad hoc}
structures and fine tuning of parameters. 
On the other hand, our goal here is to propose
a solution to the $\mu$-problem in GMSB theories which satisfies the
following criteria of naturalness: {\it i)} the different 
supersymmetry-breaking
Higgs parameters are generated by a single mechanism; {\it ii)} $\mu$
is generated at one loop, while $B_\mu$, $m_H^2$, $m_{\bar H}^2$ are
generated at two loops; {\it iii)} all new coupling constants 
are of order one;
{\it iv)} there are no new particles at the weak scale.

The paper is organized as follows: in sect.~2 we review the GMSB theories
and define the theoretical framework in which we will work. In sect.~3
we present the $\mu$-problem which we will attempt to solve in sect.~4.
Finally our results are summarized in sect.~5.

\section{The GMSB Theories} 

In this section we define the set of models which we want to study. We
first introduce an {\it observable sector} which contains the usual
quarks, leptons, and two Higgs doublets, together with their supersymmetric
partners. Next the theory has a {\it messenger sector}, formed by some
new superfields which transform under the gauge group as a real non-trivial
representation. In order to preserve gauge coupling constant unification,
we also require that the messengers form complete GUT multiplets.
Perturbativity of $\alpha_{\rm GUT}$ at the scale $M_{\rm GUT}$ implies
that we can introduce at most $n_5 ({\bf 5} + {\bf {\overline 5}})$ and
$n_{10} ({\bf 10} + {\bf {\overline{10}}})$ $SU_5$ representations with 
$n_5 \le 4$ for $n_{10}=0$ and $n_5\le 1$ for $n_{10}=1$ \cite{mur}.
It should be noticed that, in the minimal $SU_5$ model, the presence
of these new states at scales of about 100 TeV is inconsistent with
proton-decay limits and with 
$b$--$\tau$ unification \cite{murb}. However, these
constraints critically depend on the GUT model and will be dismissed
from the point onwards.

Finally the theory contains a {\it secluded sector}\footnote{We introduce
this terminology to distinguish this sector from the hidden sector of
theories where supersymmetry breaking is mediated by gravity.}.
This sector, responsible for the mechanism of supersymmetry breaking
in a gauge-invariant direction,
has tree-level couplings to the messenger sector, but not to the
observable sector. Its effect is to feed two (possibly different)
mass scales to the theory: $M$, the scale of supersymmetric-invariant
masses for the messenger superfields, and $\sqrt{F}$, the effective
scale of supersymmetry breaking or, in other words, the mass splittings
inside messenger multiplets. We will parametrize the effect of the
secluded sector by a mass term in the superpotential
\beq
W={\bar \Phi}_i M_{ij} \Phi_j ~~~~~~~~i,j=1,...,n
\label{uno}
\eeq
and by a supersymmetry-breaking term in the scalar potential
\beq
V={\bar \Phi}_{i} F_{ij} \Phi_{j}+{\rm h.c.}
\label{due} 
\eeq
Here $\Phi_i$ and ${\bar \Phi}_i$ are a generic number $n$ of messenger 
superfields transforming as the representation ${\bf r}+
{\bf {\overline r}}$ under the GUT group. 
With a standard abuse of notation, we denote the superfields, as in 
eq.~(\ref{uno}), and their scalar components, as in 
eq.~(\ref{due}), by the same symbol.
The interactions in eqs.~(\ref{uno})
and (\ref{due}) can be obtained from tree-level couplings of
the messengers $\Phi$ and $\bar \Phi$ to some superfields $X$ which
get Vacuum Expectation Values (VEV) both in their scalar components
$\langle X \rangle =M$ and their auxiliary components $\langle F_X
\rangle =F$. Supersymmetry breaking can occur dynamically, as in the models
of ref.~\cite{din}, or through perturbative interactions,
as in the O'Raifeartaigh model
\cite{oraf} described by the superpotential\footnote{In this specific
example the one-loop effective potential fixes $\langle X \rangle =0$
\cite{huq}. However one can easily extend the field content to obtain
$\langle X \rangle \ne0$.}
\beq
W=\lambda X \left({\bar \Phi}_1 \Phi_1 - m^2 \right)
+M_\Phi \left( {\bar \Phi}_1 \Phi_2 + {\bar \Phi}_2 \Phi_1 \right) ~.
\label{orafeq}
\eeq
Indeed, for $\lambda^2 m^2 <M_\Phi^2$, the vacuum of this model is such
that $\langle \Phi_i \rangle =\langle {\bar \Phi}_i \rangle =0$
($i=1,2$) and $\langle F_X \rangle \ne 0$.

Having described the necessary ingredients of the GMSB theories, we can now
proceed to compute the feeding of supersymmetry breaking into the
observable sector mass spectrum. The mass term for the messenger scalar
fields, derived from eqs.~(\ref{uno}) and (\ref{due}) is
\beq
\pmatrix{\Phi^\dagger & {\bar \Phi}}
\pmatrix{M^\dagger M & F^\dagger \cr F & MM^\dagger}
\pmatrix{\Phi \cr {\bar \Phi}^\dagger}~.
\eeq
We can now choose a basis in which the matrix $M$ is diagonal
and define
\beq 
\varphi =\frac{\Phi + {\bar \Phi}^\dagger}{\sqrt{2}}~,~~~~~~
{\bar \varphi} =\frac{{\bar \Phi} - \Phi^\dagger}{\sqrt{2}}~.
\eeq
In the new basis the scalar messenger mass term becomes
\beq
\pmatrix{\varphi^\dagger & {\bar \varphi}}
\pmatrix{M^2+\frac{F+F^\dagger}{2} & \frac{F^\dagger -F}{2}\cr 
\frac{F -F^\dagger}{2} & M^2-\frac{F+F^\dagger}{2}}
\pmatrix{\varphi \cr {\bar \varphi}^\dagger}~.
\eeq

We now want to require that there are no one-loop contributions to
squark and slepton squared masses proportional to the corresponding
hypercharge. These contributions are phenomenologically unacceptable,
since they give negative squared masses to some of the squarks and
sleptons. They arise from a one-loop contraction of the hypercharge
messenger D-term
\beq
D_\Phi =g^\prime \left( \Phi^\dagger Y_\Phi \Phi - {\bar \Phi}Y_\Phi 
{\bar \Phi}^\dagger \right)=-g^\prime 
\left( {\bar \varphi}Y_\Phi  \varphi
+\varphi^\dagger Y_\Phi  {\bar \varphi}^\dagger \right)~.
\eeq
If $F$ is Hermitian, then the two sectors $\varphi$ and $\bar \varphi$
do not mix in the mass matrix, and the D-term contributions are
not generated up to three loops\footnote{Indeed, in
the absence of the observable sector and for $F=F^\dagger$, the theory
is invariant under a parity which transforms $\varphi \to \varphi$,
${\bar \varphi}\to -{\bar \varphi}$ 
and the fermions $\psi \to \gamma_0 \psi$.}.
Therefore
we require that the secluded sector 
is such that there exists a basis where $M$ is diagonal and $F$ is Hermitian.
For instance, had we chosen different mass parameters for the terms 
${\bar \Phi}_1 \Phi_2$ and 
${\bar \Phi}_2 \Phi_1$ in eq.~(\ref{orafeq}),
there would be no cancellation of one-loop hypercharge D-term contributions
to squark and slepton masses, and the model should be discarded.

The next step is the diagonalization of the mass matrices $M^2\pm F$
for the two sectors $\varphi$ and ${\bar \varphi}$. After that, the loop
computation of the gaugino and squark masses proceeds as discussed
in ref.~\cite{gaug}.
The result is that, in the limit
where the entries of the matrix
$F$ are smaller than those of $M^2$, the gaugino
and scalar masses are\footnote{If elements of $F$ were 
larger than elements of $M^2$, the messenger squared mass matrix may 
develop negative eigenvalues. The requirement that gauge symmetry remains
unbroken in the messenger sector justifies the approximation made here.}
\beq
m_{\lambda_j}=k_j\frac{\alpha_j}{4\pi}\Lambda_G ~\left[ 
1+{\cal O}(F^2/M^4) \right]
~,~~~~~~~j=1,2,3~,
\eeq
\beq
{\widetilde m}^2=2\sum_{j=1}^3 C_jk_j\left( \frac{\alpha_j}{4\pi}\right)^2
\Lambda_S^2~\left[ 1+{\cal O}(F^2/M^4) \right]~,
\label{squak}
\eeq
where $k_1=5/3$, $k_2=k_3=1$, and $C_3=4/3$ for colour triplets, $C_2=3/4$
for weak doublets (and equal to zero otherwise), $C_1=Y^2$ 
($Y=Q-T_3$). The scales $\Lambda_G$ and $\Lambda_S$, in the limit
in which $M=M_0~\identity$, are given by
\beq
\Lambda_G=N\frac{{\rm Tr}F}{M_0}~,
\label{lamg}
\eeq
\beq
\Lambda_S=\left(N\frac{{\rm Tr}F^2}{M_0^2}\right)^{1/2}~,
\label{lams}
\eeq
where $N$ is the Casimir of the messenger GUT representation,
{\it e.g.} $N=1$ for ${\bf 5}+{\bf {\overline 5}}$ and
$N=3$ for ${\bf 10}+{\bf {\overline{10}}}$. 

All squarks and sleptons (and analogously all gauginos) receive masses
determined by a unique scale $\Lambda_S$ (or $\Lambda_G$). This universality 
is a consequence of the assumption that the secluded sector
contains only GUT singlets and of the fact that the ratio $F_{ii}/M_i$ is
not renormalized by gauge interactions.

The values of $\Lambda_G$ and $\Lambda_S$ are in general different.
If the matrix $F$ is proportional to $M$, as is the case when the
interactions in eqs.~(\ref{uno}) and (\ref{due}) originate from couplings
to a single superfield $X$, then the ratio
\beq
\Lambda_G/\Lambda_S =\sqrt{Nn}
\eeq
is directly related to the number $n$ of messenger GUT multiplets. However, in
general, the ratio $\Lambda_G/\Lambda_S$ can be either smaller or
larger than one. If $\Lambda_G\ne 0$ then $\Lambda_S \ne 0$, but the
converse is not true. For instance, in the model of eq.~(\ref{orafeq}),
$\langle X\rangle =0$ is determined by the one-loop effective potential
\cite{huq}, and $\Lambda_S \ne 0$ while $\Lambda_G = 0$, as a consequence
of an exact R-symmetry.

Finally, let us review the present bounds on the different mass scales
in the theory. The experimental limits on gluino and right-handed
selectron masses require
\beq
\Lambda_G \gappeq 16~{\rm TeV}~,~~~~~~~
\Lambda_S \gappeq 30~{\rm TeV}~.
\label{lmt}
\eeq
This imposes a model-dependent lower bound on the typical scale $M$.

It is fairly generic that the secluded sector gives rise
to an R-axion. If this is the case, one can invoke gravitational
interactions to generate a mass for the R-axion. Astrophysical constraints
can then be evaded if the typical supersymmetry breaking scale
satisfies $\sqrt{F}\lappeq 100$ TeV \cite{rax}. Upper bounds on $\sqrt{F}$
can be derived from cosmological considerations. If there is no inflation
with low reheating temperature, the constraint on the relic gravitino
density requires $\sqrt{F}< 2\times 10^3$ TeV \cite{pag}. Also if we impose
that the lightest supersymmetric particle of the observable sector
decays during the first second of the Universe, so that its decay
products cannot influence standard nucleosynthesis, then $\sqrt{F}<
10^5~{\rm TeV}~ (m_{LSP}/100~{\rm GeV})^{5/4}$.

\section{The $\mu$-Problem in GMSB Theories}
 
If the $\mu$-term is present in the superpotential in the limit of
exact supersymmetry, then naturally it can only be of the order of
the Planck scale $M_{\rm PL}$ or some other fundamental large mass
scales. We assume therefore that the $\mu$-term is forbidden in the
original superpotential but generated, together with $\bmu$, by the 
following effective operators 
\bea
\frac{1}{M}&\int &d^4\theta H\bar H X^\dagger \, ,\label{opemu}\\
\frac{1}{M^2}&\int &d^4\theta H\bar H XX^{\dagger}\, .\label{opebmu}
\eea
Here $X$ is a superfield which parametrizes the breaking of supersymmetry,
as discussed in sect.~2,
and $M$ is the 
messenger mass scale.
Replacing $X$ in eqs.~(\ref{opemu})
and (\ref{opebmu}) with the VEV of its auxiliary component $F$,
we obtain $\mu$ and $\bmu$ terms of the order of
$\Lambda$ and $\Lambda^2$ respectively, with $\Lambda=F/M$.
In theories where gravity mediates supersymmetry breaking, $M$ has to
be identified with $M_{\rm PL}$ and the operators in
eqs.~(\ref{opemu})
and (\ref{opebmu}) are present in the theory as non-renormalizable
interactions. The existence of the operator in eq.~(\ref{opemu})
requires however a non-minimal K\"ahler metric \cite{giu}.

In GMSB theories we want to obtain the operators in eqs.~(\ref{opemu})
and (\ref{opebmu}) after integrating out some heavy fields. Since the
operators in eqs.~(\ref{opemu})
and (\ref{opebmu}) break a Peccei-Quinn symmetry, they cannot be induced
by gauge interactions alone. The simplest way to generate a $\mu$-term
at the one-loop level is then
to couple the Higgs superfields to the messengers in the superpotential:
\be
W=\lambda H\Phi_1\Phi_2+\bar\lambda \bar H\bar\Phi_1\bar\Phi_2\, .
\label{couplinga}
\ee
Assuming that
a single superfield $X$ describes the supersymmetry breaking  
\be
W=X(\lambda_1 \Phi_1\bar\Phi_1+\lambda_2 \Phi_2\bar\Phi_2)\, ,
\label{couplingb}
\ee
the diagram of fig.~1a gives
\be
\mu=\frac{\lambda\bar\lambda}{16\pi^2}\Lambda ~ f(\lambda_1/\lambda_2 )
~\left[ 1+{\cal O}(F^2/M^4) \right]
\, ,
\label{muloop}
\ee
$f(x)=(x\ln x^2)/(1-x^2)$.
However, the couplings in eq.~(\ref{couplinga}) also generate
the diagram of fig.~1b,  which contributes to  $\bmu$
at the one-loop level:
\be
\bmu=\frac{\lambda\bar\lambda}{16\pi^2}\Lambda^2 ~f(\lambda_1/\lambda_2 )
~\left[ 1+{\cal O}(F^2/M^4) \right]
\, .\label{bmuloop}
\ee
Finally the diagram of fig.~1c generates the soft-breaking masses
$m_H^2$ and $m_{\bar H}^2$. Unexpectedly, the leading order contribution
here cancels and the scalar Higgs masses are generated only at higher order
$\sim \frac{1}{16\pi^2}F^4/M^6$. 
However the cancellation is valid only in the simple 
case of eq.~(\ref{couplingb}) in which the ratios $\Lambda_i=F_{ii}/M_i$ 
($i=1,2$) for the two messengers are the same.
If this is not the
case then
\be
m_{H}^2= \frac{\lambda^2}{16\pi^2}(\Lambda_1
-\Lambda_2 )^2~g(\lambda_1/\lambda_2 )
~\left[ 1+{\cal O}(F^2/M^4) \right]
\, ,
\ee
and similarly for $m_{\bar H}^2$ with $\lambda$ replaced by $\bar \lambda$;
here $g(x)=x^2[(1+x^2)\ln x^2+2(1-x^2)]/(1-x^2)^3$.

{}From eq.~(\ref{muloop}) and 
(\ref{bmuloop}) we obtain
\be
\bmu=\mu\Lambda\, .\label{problematic}
\ee
This problematic relation is the expression of the $\mu$-problem in
GMSB theories. It is just a consequence of generating both $\mu$
and $\bmu$ at the one-loop level through the same interactions.
If the dominant contributions to $m_H^2$ and $m_{\bar H}^2$ come from
two-loop gauge and three-loop stop contributions \cite{din},
\bea
m^2_{ H}&=&\frac{3}{2}\left(\frac{\alpha_2}{4\pi}\right)^2\Lambda^2\, ,
\nonumber\\
m^2_{\bar H}&=&m^2_{H}\left[1-\frac{4h_t^2}{3\pi^2}\left(\frac{\alpha_3}
{\alpha_2}\right)^2\ln \left(\frac{\pi}{\alpha_3}\right)\right]\, ,
\label{stop}
\eea
then eq.~(\ref{problematic}) is actually
inconsistent with electroweak symmetry breaking. 
This is true because the condition for the stability of the Higgs potential,
 $2|\bmu|<m^2_{H}+m^2_{\bar H}+2\mu^2$, cannot be satisfied when $\mu$
is determined by electroweak symmetry breaking
\be
|\mu |^2= \frac{m^2_{H}-m^2_{\bar H}\tan^2\beta}{\tan^2\beta-1}-
\frac{m^2_Z}{2}\, ,
\label{break}
\ee
with $\tan\beta=\langle \bar H\rangle/\langle H\rangle$.
Extra contributions
to $m_H^2$ and $m_{\bar H}^2$ can allow the electroweak symmetry
breaking, but only at the price of introducing a considerable
fine tuning among parameters.

The alternative to coupling each Higgs superfield separately to the 
messengers, as done in eq.~(\ref{couplinga}), is to couple the Higgs
bilinear $H\bar H$ to a heavy singlet superfield $S$. Of course, for
the mechanism to work, $S$ cannot develop a VEV in the supersymmetric limit.
Now, possibly through a coupling to a different singlet $N$ which has
tree-level non-zero VEV, a $\mu$-term can be induced at the one-loop
level through the graph in fig.~2a \cite{graphs}. 
However with an extra insertion of
the spurion superfield $X$, the same interaction gives rise to a $\bmu$
contribution through the graph in fig.~2b and we recover the
unwanted relation of eq.~(\ref{problematic}).

Nevertheless, this is not always the case. Suppose that
we couple $N$ to the  messengers in the same way as we coupled $H$ to
the messengers in
eq.~(\ref{couplinga}). From the previous example, we know that the
diagram of fig.~1c suffers from a cancellation of the leading contribution in
the case of the minimal secluded sector given in
eq.~(\ref{couplingb}). This means that also the diagram in fig.~2b 
vanishes at the leading order.
Extra $X$ and $X^\dagger$ or 
extra $N$ and $N^\dagger$ insertions in the loop
of fig.~2b  can generate a non-zero $\bmu$ but this will
be suppressed by terms $\calo(F^2/M^4)$ or 
$\calo(\langle N\rangle^2/M^2)$, if $M$ is 
larger than the other  scales. However we do not consider this mechanism
entirely satisfactory, since it relies on an accidental cancellation
occurring only for a very specific supersymmetry-breaking structure.

Finally we mention that in the literature the possibility of
adding extra light Higgs superfields in GMSB theories
has also been considered \cite{din}. 
If a light singlet $S$ is present with a 
superpotential 
\be
W=\lambda H\bar HS+\lambda'S^3\, ,
\ee
then $\mu$ and $\bmu$ are generated whenever $\langle S\rangle$
and $\langle F_S\rangle$ are non-zero.
However, in GMSB theories, 
trilinears, bilinears and the soft-mass of $S$ are suppressed with 
respect to the $H$ and $\bar H$ soft masses, and 
a non-zero VEV for $S$ requires an appreciable fine-tuning \cite{din}.
This problem can be overcome but its solution may require  
additional quark superfields
coupled to 
$S$ in order to induce a large soft mass $m_S^2$ \cite{din}.

\section{A Natural Solution to the $\mu$-Problem in GMSB Theories}

\subsection{The Mechanism}
We will describe here a mechanism which satisfies the criteria
$(i)-(iv)$ given in sect.~1. We consider 
the possibility that $\mu$,
instead of being generated by the operator (\ref{opemu}), 
 arises from the operator
\be
\int d^4\theta H\bar H D^2\left[X^\dagger X\right]\, .
\label{qqq}
\ee
Here $D_\alpha$ is the supersymmetric covariant derivative. 
This operator can be generated from the diagram of fig.~3. 
The crucial point is  that a $\bmu$-term cannot be induced 
from such diagram even if we added
extra $X$ and $X^\dagger$ insertions in the loop of fig.~3. This is because a 
$D^2$ acting on any function of $X$ and $X^\dagger$ always produces
an antichiral superfield. 

Our mechanism requires at least
two singlets, $S$ and 
$N$, such that only $S$  couples at tree-level to 
$H\bar H$ and to the messengers.
 We forbid the coupling of $N$ to 
the messengers and a mass term $S^2$ in the superpotential
to guarantee that the one-loop diagram
of fig.~2b does not exist.
The diagram of fig.~3  induces the operator
\be
\frac{1}{16\pi^2M^2M^2_N}
\int d^4\theta H\bar H D^2\left[X^\dagger X\right]\, ,
\label{newmu}
\ee
where  $M_N$ is the 
 mass parameter in the superpotential term $M_NNS$.
Notice that the internal line in fig.~3 is an $\langle SS^\dagger\rangle$
propagator which, at small momenta, behaves like ${\bar D}^2/M_N^2$
\cite{graphs}.
Eq.~(\ref{newmu}) leads to a $\mu$ parameter 
\be
\mu\sim\frac{1}{16\pi^2}\frac{|F|^2}{MM^2_N}\sim
\frac{1}{16\pi^2}\Lambda\, ,
\label{ping}
\ee
for $M_N=\calo(\sqrt{F})$.

$\bmu$ can be induced at the two-loop level
if the superpotential contains a coupling of the form $N^2S$.
One contribution comes from a diagram analogous to the one shown in fig.~2b,
but where the effective coupling between $N$ and $X$ arises at two loops,
as shown in fig.~4. Another contribution comes from the diagram of
fig.~5, which induces the effective operator
\be
\frac{1}{(16\pi^2)^2M^4M^4_N}\int d^4\theta H\bar H X^{\dagger}X\bar D^2
D^2\left[X^\dagger X\right] \, .
\label{newbmu}
\ee
Both contributions generate a $\bmu$ parameter of the correct magnitude
\be
\bmu\sim\left(\frac{1}{16\pi^2}\right)^2\Lambda^2 \, .
\label{pong}
\ee
In addition to the gauge-induced contributions given in eq.~(\ref{squak}),
the soft masses $m_H^2$ and $m_{\bar H}^2$ are also generated at two loops by
the diagrams in fig.~4.

\subsection{A Model}
Let us now see how to realize this mechanism in an explicit example.
Consider the superpotential
\beq
W=S(\lambda_1 H \bar H +\frac{\lambda_2}{2}N^2+\lambda \Phi \bar \Phi
-M_N^2) ~,
\label{spot}
\eeq
together with a secluded sector which, as described in sect.~2, generates 
a supersymmetric mass $M$ and a supersymmetry-breaking squared mass
$F$ for the messengers $\Phi$ and $\bar\Phi$. For simplicity we assume
that the messengers belong to a single
${\bf 5}+{\bf \overline 5}$ $SU_5$ representation. 

The expression in eq.~(\ref{spot}) can be guaranteed by gauge symmetry,
a discrete parity of the superfield N, and an R-symmetry. We believe
that eq.~(\ref{spot}) describes the simplest example in which our
mechanism is operative. The tree-level coupling of the singlet $S$ with
the messengers in eq.~(\ref{spot}) is crucial since it induces, through
one-loop radiative corrections, a tadpole for $S$, which generates
$\langle S\rangle \ne 0$ and the desired $\mu$-term. For a tadpole
diagram to exist, it is also essential that $S$ does not transform 
under any discrete or continuous symmetry left unbroken by the 
secluded sector. In the case of eq.~(\ref{spot}), the R-symmetry under
which $S$ transform non-trivially is certainly broken (possibly
spontaneously) in the secluded
sector, or else no gaugino mass is generated.

The tree-level potential is
\begin{eqnarray}
V &=& \left | \lambda_1 H\bar H + \frac{\lambda_2}{2} N^2+ 
\lambda \Phi \bar{\Phi}- M_N^2 \right | ^2 + 
|S|^2 \left[ \lambda_1^2 (|H|^2 +|\bar H|^2) + 
\lambda_2^2 |N|^2 \right] \nonumber\\
&+& |S\lambda + M|^2 \left ( |\Phi|^2 + |\bar{\Phi}|^2 \right )
+ \left( F\Phi\bar{\Phi} + \rm{h.c.}\right) ~.
\label{poten}
\end{eqnarray}
For $M^2>F$, a minimum of the potential is at
\beq
\langle \lambda_1 H\bar H +\frac{\lambda_2}{2}N^2\rangle =M_N^2~,
\label{vev}
\eeq
and all other VEVs equal to zero. The two Higgs doublets are massless at 
tree level and can be viewed as zero modes of the flat direction determined
by eq.~(\ref{vev}). 

One-loop corrections induce a tadpole for the scalar component of
$S$
\beq
V=\frac{5\lambda}{16\pi^2}\frac{F^2}{M}~S+{\rm h.c.}~,
\label{tad}
\eeq
which forces $\langle S\rangle \ne 0$. 
Inspection of the
effective potential shows that there are no runaway directions at
large values of $S$,
as discussed in the appendix. 
Radiative corrections also remove
the degeneracy of the vacuum in eq.~(\ref{vev}). 
An important role is played by
the two-loop contributions to the
soft masses of $N$, $H$, and $\bar H$,
obtained from the diagrams of fig.~4 which, for $M>M_N$, give:
\beq
m_N^2=10\left( \frac{\lambda \lambda_2}{16\pi^2}\right)^2\frac{F^2}{M^2}~,
\eeq 
\beq
\delta m_H^2=\delta m_{\bar H}^2=10
\left( \frac{\lambda \lambda_1}{16\pi^2}\right)^2\frac{F^2}{M^2}~.
\label{hb}
\eeq
The mass parameters $m_H^2$ and $m_{\bar H}^2$ are then obtained by
adding eq.~(\ref{hb}) to the gauge and stop contributions given in
eq.~(\ref{stop}).
Including the contributions in eqs.~(\ref{tad})--(\ref{hb})
to the potential in eq.~(\ref{poten}) and
proceeding in the minimization, we find that 
the VEV in eq.~(\ref{vev}) predominantly lies in the $N$ direction and
$\langle S\rangle$ is determined to be
\beq
\langle S \rangle \simeq -\frac{5}{32\pi^2}\frac{\lambda}{\lambda_2}
\frac{F^2}{M_N^2M}~,
\eeq
as long as $\lambda_2(\lambda_1^2\langle S\rangle^2+m_H^2) >
\lambda_1(\lambda_2^2\langle S\rangle^2+m_N^2) $.
The two Higgs doublets are the only superfields to remain light. They
have the usual low-energy supersymmetry potential with the parameters
$\mu$ and $\bmu$ given by
\beq
\mu =\lambda_1\langle S\rangle 
=-\frac{5}{32\pi^2}\frac{\lambda\lambda_1}{\lambda_2}\frac{F}{M_N^2}
\Lambda ~,
\label{zip}
\eeq
\beq
\bmu = \lambda_1\langle F_S \rangle =-\frac{10\lambda^2\lambda_1
\lambda_2}{(16\pi^2)^2}
\left( 1+\frac{5F^2}{8\lambda_2^2M_N^4}\right)\Lambda^2 ~.
\label{zap}
\eeq
Equations (\ref{zip}) and (\ref{zap}) confirm in this specific model
the estimates, based on general arguments, given in eqs.~(\ref{ping})
and (\ref{pong}).

\subsection{Pseudo-Goldstone Boson Interpretation}

The mechanism that generates
$\bmu\sim\mu^2$, instead of
$\bmu\sim\mu\Lambda$, has a pseudo-Goldstone boson interpretation.
Let us modify the previous model by introducing a new gauge singlet
$\bar N$ and by replacing eq.~(\ref{spot}) with
\be
W=S(\lambda_1 H \bar H +{\lambda_2}N\bar N+\lambda \Phi \bar \Phi
-M_N^2) ~.
\ee
The results of sect.~4.3 are essentially unaffected. However here,
in the limit $\lambda_1 = \lambda_2$, the superpotential 
has a U(3) symmetry under which $\Sigma\equiv (H,N)$ and
 $\bar{\Sigma}\equiv (\bar H, \bar N)$ 
transform as a triplet and an anti-triplet.
In the supersymmetric limit,
the VEVs of $N$ and $\bar N$ break the
$U(3)$ spontaneously to U(2) and
the two Higgs doublets are identified with the corresponding 
Goldstone bosons\footnote{The idea of interpreting the Higgs doublets
at pseudo-Goldstone bosons of some large global symmetry 
of the superpotential has been introduced
in refs.~\cite{pgb,geo}.}. Actually they are only pseudo-Goldstone
bosons since they get non-zero masses as soon as gauge 
and Yukawa interactions
are switched-on\footnote{The theory contains also an exact Goldstone bosons,
which corresponds to the spontaneous breaking of the abelian symmetry
carried by the $N$ and $\bar N$ superfields. However this Goldstone boson
has no coupling to ordinary matter.}.
Nevertheless,
at the one-loop level,
the relevant part of the effective potential is still U(3)-invariant
and  one  combination of the two Higgs doublets
$({H + \bar H^\dagger)/\sqrt{2}}$, remains exactly massless.
Indeed, at one loop, 
the determinant of the Higgs squared-mass matrix is zero, and
\be
\bmu=|\mu|^2\, ,\label{pgbeq}
\ee
a general property of models in which the Higgs particles are 
pseudo-Goldstone bosons \cite{fine}. Soft masses for $H$ and $\bar H$
are generated at two loops. The contributions from the graphs in
fig.~4 preserve the $U(3)$ invariance, but gauge contributions violate
the symmetry and the determinant 
of the Higgs squared-mass matrix no longer vanishes.
Nevertheless, we are still guaranteed to obtain a $\mu$ and $\bmu$ of
the correct magnitude, since eq.~(\ref{pgbeq}) is spoiled only by
two-loop effects. If we now allow $\lambda_1 \ne \lambda_2$, we will
modify eq.~(\ref{pgbeq}) but not the property $\bmu \sim \mu^2$.
This provides an explanation, alternative to the one given in
sect.~4.1 in terms of effective operators, of the reason
why our mechanism can work.

Notice 
that, although for $\lambda_1\not= \lambda_2$ the U(3) is no longer
a symmetry of the superpotential, the vacuum $\langle \lambda_1
H\bar H +\lambda_2 N\bar N\rangle =M_N^2$ still has a compact
degeneracy isomorphic to $U(3)/U(2)$.
This degeneracy of the tree-level vacuum in our model,
need not be necessarily an accident of the field
structure of the low-energy sector. It may result from an
exact U(3), or even SU(6), gauge  symmetry spontaneously broken
at some high scale in a heavy sector that does not couple with
our fields in the superpotential. To be specific, imagine the embedding
of our theory in the gauge SU(3)$_C\otimes$SU(3)$_L\otimes$U(1)
model with $\Sigma$ and $\bar{\Sigma}$ transforming
as triplets
and anti-triplets of SU(3)$_L$ respectively. The assignment of quarks and
leptons can be  easily fixed if we think of
SU(3)$_C\otimes $SU(3)$_L\otimes $U(1) as a maximal subgroup of
SU(6), with quarks and
leptons transforming as $ {\bf 15} + {\bf \bar 6} + {\bf \bar 6}$ of SU(6).
Suppose that at some high scale there is another sector
(\eg\ , an extra triplet-antitriplet
pair $\Sigma',\bar{\Sigma}'$) 
which does not communicate with
$\Sigma,\bar{\Sigma}$ in the superpotential.
In such a case the
Higgs superpotential has an accidental  U(3)$_L\otimes $U(3)$_L$
approximate symmetry corresponding to the independent global transformation
of $\Sigma,\bar\Sigma$ and $\Sigma'\bar\Sigma'$ \cite{geo}.
The crucial point is that 
a residual global  U(3)$_L$ symmetry is left 
at low energy,
if the breaking of the gauge symmetry
SU(3)$_L\otimes $U(1)$ \rightarrow $SU(2)$_L\otimes $U(1)$_Y$
is  induced only by the VEV of $\Sigma'\bar\Sigma'$ at the high-energy scale.

\section{Conclusions}
 
Let us summarize our results. In sect.~2 we have presented the GMSB theories
with a general structure of messenger superfields. We have given the
consistency conditions for not generating dangerous negative squark
squared masses, and given the general expressions of squark, slepton, and
gluino masses. Within our approximations, the ignorance on the messenger
sector can be parametrized by the two mass scales $\Lambda_G$ and $\Lambda_S$.

We have shown in sect.~3 that GMSB theories suffer from a $\mu$-problem which
has a different aspect than the $\mu$-problem in supergravity. 
The difficulty here is that once $\mu$ is generated by a one-loop diagram,
$\bmu$ also arises at the same loop level; this leads to the problematic
relation in eq.~(\ref{problematic}).

In sect.~4 we have considered a mechanism which evades this generic
problem. If $\mu$ is generated by the effective operator (\ref{qqq}),
then $\bmu$ is not necessarily induced at the same loop order. We have
presented a simple model in which this idea is realized explicitly.
This mechanism can naturally lead to an interpretation of the Higgs
doublets as pseudo-Goldstone bosons of an approximate global symmetry.

We would like to thank R.~Barbieri,
S.~Dimopoulos, and S.~Raby for very useful discussions.
\vskip .2in

\noindent{\Large{\bf Appendix}}
\vskip .2in
\noindent
In this appendix we want to show that the potential of the model
considered in sect.~4.2
does not have runaway directions in the large $S$ region.
First note that for sufficiently large values $S  > S_c$,
such that  $|S_c|^2 > M_N^2/\lambda_1,M_N^2/\lambda_2$ and 
$|S_c\lambda + M|^2 > |\lambda M_N^2 - F|$, the minimum 
of all fields (other than $S$) at a given
fixed $S$ is at zero. So, for $S > S_c$
the tree-level potential is flat in the $S$ direction and has
a constant value $V (S > S_c) = M_N^4$.  
It is therefore essential to look at the one-loop corrections to
the effective potential 
\begin{equation}
V_{eff}=
{1 \over 64 \pi^2} 
 (-1)^F \rm{Tr}~ {\cal M}^4 ~\ln{{\cal M}^2 \over Q^2}~.
\end{equation}
For $S > S_c$, all superfields interacting
with $S$ in the superpotential suffer from tree-level mass
splittings caused by the non-zero $F_S = -M_N^2$, and therefore contribute to
$V_{eff}$. Evaluating and adding different contributions we get the
asymptotic behaviour
\begin{eqnarray}
V_{eff}|_{|S|\to \infty} &=&\frac{1}{16\pi^2}
\left[2\lambda_1^2M_N^4\ln(\lambda_1^2|S|^2)+
\frac{\lambda_2^2}{2}M_N^4\ln(\lambda_2^2|S|^2)\right. \nonumber \\
&+&\left. 5(\lambda M_N^2-F)^2\ln |\lambda S+M|^2\right] ~,
\end{eqnarray}
which shows that $V_{eff}$ grows logarithmically at large $S$.

{}From the effective potential it is also easy to see that quantum
corrections destabilize the tree-level vacuum $\langle S\rangle =0$.
For small $S$, none of the states $S,\bar H,H,\bar N,N$
contribute to the effective potential,
since there is no tree level mass splitting inside
these multiplets. Thus, the only states that contribute
are the messengers $\Phi, \bar{\Phi}$. At tree level these states have
$S$-dependent supersymmetric mass-squared $|\lambda S + M|^2$ and
mass-splittings $ \pm (F - \lambda\lambda_i |S|^2) $
between the real and imaginary parts of their scalar
components.
The $S$-dependence of this splitting comes from the $F_S$-term which, 
for $S \neq 0$, is equal to $ - \lambda_i|S|^2$ where $i = 1$
(or 2) if $|\lambda_2| > |\lambda_1|$ (or 
$|\lambda_2| < |\lambda_1|$). Evaluating these terms, we get the
following result
\begin{equation}
\left ( {\partial V_{eff} \over \partial S} \right )_{S=0} =
{5 \over 16 \pi^2}  \lambda M \sum_\pm (M^2 \pm F)\ln
\left ( 1 \pm {F \over M^2} \right )~.
\nonumber
\end{equation}
which, for $M>>F$ 
corresponds to the tadpole contribution given in eq.~(\ref{tad}).

\def\ijmp#1#2#3{{\it Int. Jour. Mod. Phys. }{\bf #1~}(19#2)~#3}
\def\pl#1#2#3{{\it Phys. Lett. }{\bf B#1~}(19#2)~#3}
\def\zp#1#2#3{{\it Z. Phys. }{\bf C#1~}(19#2)~#3}
\def\prl#1#2#3{{\it Phys. Rev. Lett. }{\bf #1~}(19#2)~#3}
\def\rmp#1#2#3{{\it Rev. Mod. Phys. }{\bf #1~}(19#2)~#3}
\def\prep#1#2#3{{\it Phys. Rep. }{\bf #1~}(19#2)~#3}
\def\pr#1#2#3{{\it Phys. Rev. }{\bf D#1~}(19#2)~#3}
\def\np#1#2#3{{\it Nucl. Phys. }{\bf B#1~}(19#2)~#3}
\def\mpl#1#2#3{{\it Mod. Phys. Lett. }{\bf #1~}(19#2)~#3}
\def\arnps#1#2#3{{\it Annu. Rev. Nucl. Part. Sci. }{\bf
#1~}(19#2)~#3}
\def\sjnp#1#2#3{{\it Sov. J. Nucl. Phys. }{\bf #1~}(19#2)~#3}
\def\jetp#1#2#3{{\it JETP Lett. }{\bf #1~}(19#2)~#3}
\def\app#1#2#3{{\it Acta Phys. Polon. }{\bf #1~}(19#2)~#3}
\def\rnc#1#2#3{{\it Riv. Nuovo Cim. }{\bf #1~}(19#2)~#3}
\def\ap#1#2#3{{\it Ann. Phys. }{\bf #1~}(19#2)~#3}
\def\ptp#1#2#3{{\it Prog. Theor. Phys. }{\bf #1~}(19#2)~#3}

\vskip .2in

\noindent{\Large{\bf Figure Captions}}
\vskip .2in
\noindent{\bf Fig.~1:} {Superfield Feynman diagrams for generating
one-loop contributions to (a)
$\mu$, (b) $\bmu$, and (c) $m_H^2$, $m_{\bar H}^2$.}
\vskip .2in
\noindent{\bf Fig.~2:} {Superfield Feynman diagrams for generating
one-loop contributions to (a)
$\mu$ and (b) $\bmu$. The encircled
cross denotes the VEV of the $N$ scalar component.}
\vskip .2in
\noindent{\bf Fig.~3:} {Superfield Feynman diagram for generating a
one-loop contribution to
$\mu$.}
\vskip .2in
\noindent{\bf Fig.~4:} {Superfield Feynman diagrams for generating
two-loop contributions to $m_N^2$, $m_H^2$, and $m_{\bar H}^2$.}
\vskip .2in
\noindent{\bf Fig.~5:} {Superfield Feynman diagram for generating a
two-loop contribution to $\bmu$.}

\vfill\eject

\begin{figure}
\hglue1.5cm
\epsfig{figure=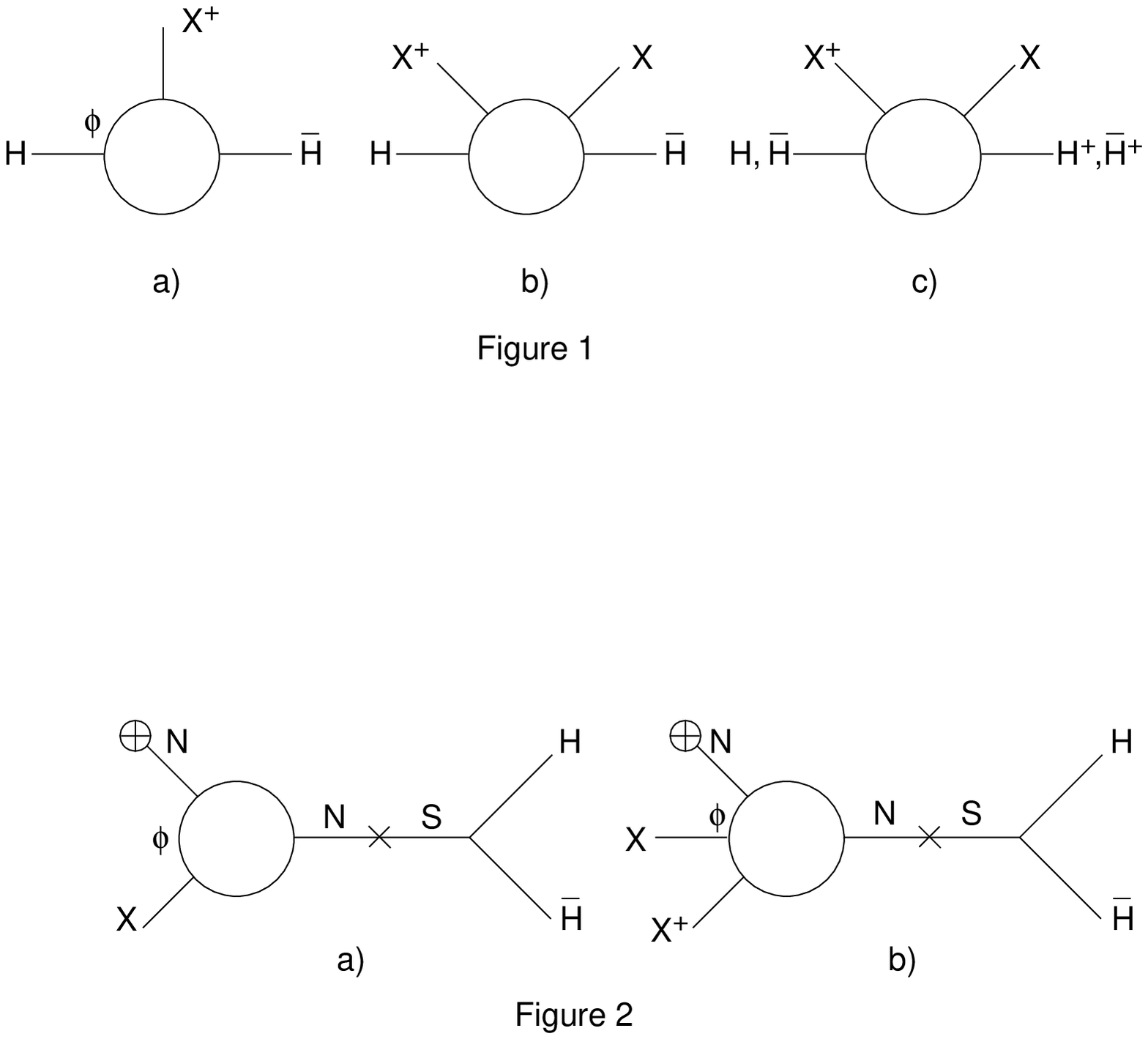,width=15cm}
\end{figure}

\begin{figure}
\hglue4.0cm
\epsfig{figure=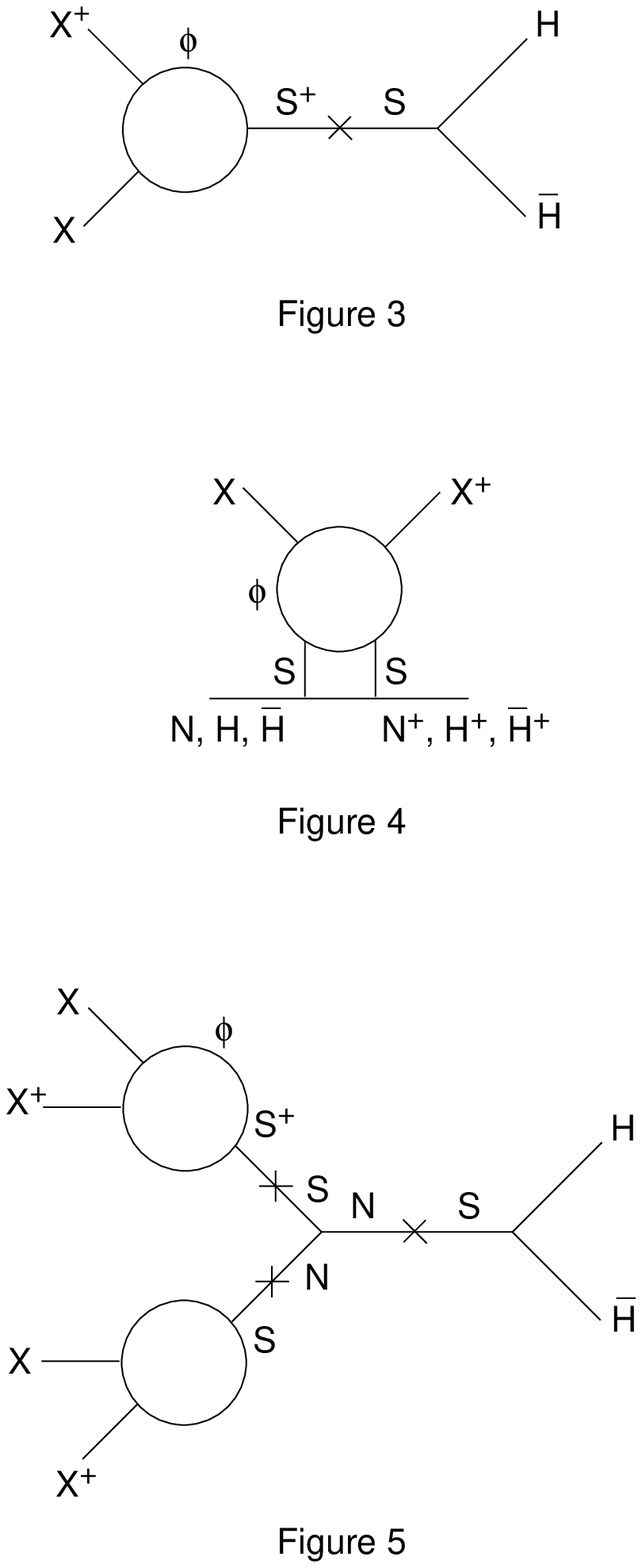,width=9cm}
\end{figure}
\end{document}